\documentstyle[12pt,epsfig]{article}
\newcommand{\simlt}  {\raisebox{-.6ex}{$\stackrel{\textstyle <}{\sim}$}}

\begin{document}
\begin{flushright}
hep-ph/0302025 \\
RAL-TR-2002-025 \\
9 Feb 2003 \\
\end{flushright}
\vspace{2 mm}
\begin{center}
{\Large
Permutation Symmetry,
Tri-Bimaximal Neutrino Mixing
and}
\end{center}
\vspace{-7mm}
\begin{center}
{\Large 
the $S3$ Group Characters
}
\end{center}
\vspace{3mm}
\begin{center}
{P. F. Harrison\\
Physics Department, Queen Mary University of London,\\
Mile End Rd. London E1 4NS. UK \footnotemark[1]}
\end{center}
\begin{center}
{and}
\end{center}
\begin{center}
{W. G. Scott\\
Rutherford Appleton Laboratory,\\
Chilton, Didcot, Oxon OX11 0QX. UK \footnotemark[2]}
\end{center}
\vspace{3mm}
\begin{abstract}
\baselineskip 0.6cm
\noindent
We postulate 
that the neutrino mass matrix
in the lepton flavour basis
is an $S3$ group matrix
in the natural representation of $S3$.
This immediately requires
one neutrino to be trimaximally mixed,
as suggested 
by the solar neutrino data.
We go on to postulate that 
the charged-lepton mass matrix
in the neutrino mass-basis 
is an $S3$ class matrix
in the natural representation 
of the $S3$ class-algebra,
leading to exact tri-bimaximal mixing,
which is compatible with data overall.
The tri-bimaximal mixing matrix is seen to be
closely related to the $S3$ character table,
and is properly the $S3 \supset S2$ 
table of induction
coefficients, where
the $S2$ corresponds to symmetry
under $\mu-\tau$ interchange
in the lepton flavour basis.

\end{abstract}
\begin{center}
{\em To be published in Physics Letters B}
\end{center}
\footnotetext[1]{E-mail:p.f.harrison@qmul.ac.uk}
\footnotetext[2]{E-mail:w.g.scott@rl.ac.uk}
\newpage
\baselineskip 0.6cm

\noindent {\bf 1. Tri-Bimaximal Mixing: 
A Problem for Democracy} \vspace{2mm}

\noindent 
We previously emphasised \cite{hps5} 
the phenomenological promise
of the so-called \cite{hsym} \cite{buch} \cite{xing} \cite{azee}
tri-bimaximal hypothesis,
defined by the lepton mixing matrix \cite{hps5} \cite{hsym}:
\begin{eqnarray}
     \matrix{  \hspace{0.4cm} \nu_1 \hspace{0.3cm}
               & \hspace{0.2cm} \nu_2 \hspace{0.3cm}
               & \hspace{0.2cm} \nu_3  \hspace{0.4cm} }
                                      \hspace{0.2cm} \nonumber \\
\hspace{1.0cm}U \hspace{0.3cm} = \hspace{0.3cm}
\matrix{ e \hspace{0.2cm} \cr
         \mu \hspace{0.2cm} \cr
         \tau \hspace{0.2cm} }
\left( \matrix{ \sqrt{\frac{2}{3}}  &
                     \frac{1}{\sqrt{3}} &
                                   0  \cr
 -\frac{1}{\sqrt{6}}  &
          \frac{1}{\sqrt{3}} &
 -\frac{1}{\sqrt{2}}   \cr
      \hspace{2mm}
 -\frac{1}{\sqrt{6}} \hspace{2mm} &
         \hspace{2mm}
            \frac{1}{\sqrt{3}} \hspace{2mm} &
                                 \frac{1}{\sqrt{2}}
                                       \hspace{2mm} \cr } \right)
\label{tb0}
\end{eqnarray}
Since then, several new experimental results,
especially the SNO \cite{sno}
flux-independent solar neutrino result,
updated measurements
from SAGE \cite{sage} and GALLEX/GNO \cite{gno}
and the recent KAMLAND \cite{kaml} reactor result, 
have considerably strengthened the case.
Given that we already know that
$|U_{e3}|^2$ $\simlt$ $0.03$ 
from reactors \cite{react},
one might even say
that the evidence for 
the solar neutrino being trimaximally mixed, 
in particular $|U_{e2}|^2$ 
\mbox{$\simeq  0.34 \pm$ \raisebox{0.9ex}{0.05} 
\raisebox{-0.9ex}{\hspace{-8.8mm}0.04}} 
from SNO \cite{sno},
is now better than the evidence for 
the atmospheric neutrino to be bimaximally mixed,
cf.\ $|U_{\mu 3}|^2 \simeq 0.50 \pm 0.11$ 
in SUPER-K \cite{skatm}.
 
Tri-bimaximal mixing is sometimes incorrectly
linked (eg.\ Ref.~\cite{altrev})
with the original rank-one democratic mass matrix
(defined by all 
mass-matrix elements equal \cite{harari}).
In fact, as we will see,
the mass matrices associated with tri-bimaximal mixing
are very far from the democratic form (Sections~4-5 below).
It is true that the democratic mass matrix  
has always one trimaximal eigenvector
($1/\sqrt{3}$, $1/\sqrt{3}$, $1/\sqrt{3}$)
\cite{fxing} \cite{simp} \cite{hps2},
but the problem is that
it is always the {\em heaviest} mass-eigenstate
(or in fact, more generally,
the {\em non-degenerate} mass-eigenstate, see below)
which ends up trimaximally mixed,
ie.\ {\em not}
what is needed phenomenologically
(cf.\ Eq.~\ref{tb0}).
In particular,
the democratic neutrino mass matrix 
can {\em in no way}\, be taken 
as a zeroth-order approximation 
for a mixing scenario
where it is the solar neutrino
(normally the intermediate mass neutrino $\nu_2$)
which is trimaximally mixed,
as in the case of tri-bimaximal mixing Eq.~\ref{tb0}. 
 
$S3$ symmetry
remains interesting.
It has 
been remarked \cite{branco}
that the $S3$ invariance
of the democratic matrix 
is unbroken
under the addition of any multiple
of the identity matrix
to the purely democratic form
(if the democratic 
component has negative sign
then an inverted hierarchy results,
still with full $S3$ invariance).
Indeed, taking any
polynomial function of a matrix
preserves all the symmetries,
and in general gives any eigenvalues 
associated with the 
original eigenvectors in any order
(the Vandermonde matrix \cite{vdm}  
formed from the original eigenvalues
provides the transformation 
between the required polynomial coefficients 
and the desired eigenvalues).
The problem 
with the democratic mass matrix here
is that (up to a factor) it is `idempotent',
ie.\ its square is not an independent matrix
and so we have in effect
only two polynomial coefficients available
(with two degenerate eigenvalues,
not only are the two corresponding eigenvectors 
undefined,
but the Vandermonde matrix has no inverse).

In this paper,
we `solve' the problem indicated above, 
by suggesting that
the democratic mass matrix be dropped,
in favour of 
an $S3$ `group matrix' \cite{little}
(more precisely, see below, 
by an element of the $S3$ group algebra,
in the sense of representation \mbox{theory)}
in the natural representation of the $S3$ group.
Later in the paper
we extend our argument
to give a new and succinct prescription 
leading to tri-bimaximal mixing itself. \\

\noindent {\bf 2. Development of Our Approach 
with a Familiar Example} \vspace{2mm}

\noindent
Although
the original 
trimaximal mixing scheme \cite{hs1} \cite{hps1}
seems now essentially ruled-out
by experiment,
it cannot be denied that
trimaximal mixing
occupied a special place
in the space
of all possible mixings.
In the briefest terms,
one had only to require
that the neutrino mass matrix
in the lepton flavour basis
(or the charged-lepton mass matrix
in the neutrino mass basis)
was a $C3$ group matrix
in the natural representation of $C3$ (see below),
and the lepton mixing matrix
was completely determined 
to be the trimaximal mixing matrix,
identically the $C3$ group character table
(see Appendix A)
up to an overall normalisation factor $1/\sqrt{3}$ \cite{char}. 
($C3$ here is the cyclic group on three objects,
while $S3$ 
above is the corresponding symmetric group). 

Explicitly,
a $C3$ group matrix \cite{little} 
is just an element 
of the $C3$ group algebra,
ie.\ an arbitrary linear combination
of the three $C3$ group elements,
with arbitrary (complex) coefficients.
(In the context of 
group representation theory,
multiplication of group elements
by scalars and addition of group elements
are understood in the obvious way,
within a given matrix representation).
In the natural representation of $C3$
(using cycle notation 
and taking $I$ to denote the identity)
the $C3$ group elements may be written:
\begin{eqnarray}
I = \left( 
\matrix{ 1 & 0 & 0 \cr
         0 & 1 & 0 \cr
         0 & 0 & 1 } \right) \hspace{3.5cm} \label{ident} \\  
P(\alpha \beta \gamma ) = \left( 
\matrix{ 0 & 1 & 0 \cr
         0 & 0 & 1 \cr
         1 & 0 & 0 } \right)  
\hspace{1.0cm}
P(\gamma \beta \alpha ) = \left( 
\matrix{ 0 & 0 & 1 \cr
         1 & 0 & 0 \cr
         0 & 1 & 0 } \right).  \hspace{1.0cm} \label{even}
\end{eqnarray}
From the  physics point of view,
if we restrict consideration
to left-handed fields only,
we may (as usual) take our mass matrices 
to be hermitian ($M \rightarrow M^2 := MM^{\dagger}$),
whereby there is nothing to be gained
by considering $C3$ group matrices
which are more general than 
the hermitian combination:
\begin{equation}
M^2=aI+bP(\alpha \beta \gamma )+b^*P(\gamma \beta \alpha ) \label{circ}
\end{equation}
where $a$ is real, $b$ is complex, 
and $b^*$ 
is the complex conjugate of $b$
(ie.\ making $a$ complex and replacing $b^*$ 
by an arbitrary complex parameter $c$,
simply yields the same form Eq.~\ref{circ},
on taking the hermitian square, 
$M \rightarrow MM^{\dagger}$).

We see immediately 
that Eq.~\ref{circ}
is just the familiar $3 \times 3$ 
circulant mass matrix \cite{hs1}:
\begin{equation}
M^2 = 
\left( \matrix{ a & b & b^* \cr
                b^* & a & b \cr
                b & b^* & a }  \right) \label{abc}
\end{equation}
invariant under cyclic ($C3$)
permutations of the generation indicies.
Diagonalising Eq.~\ref{abc} is equivalent 
to reducing the $C3$ group algebra 
to independent idempotents \cite{little}
(the projection operators of Ref.~\cite{hps4})
and leads directly 
to trimaximal mixing \cite{hps1}:
\begin{equation}
U = \frac{1}{\sqrt{3}}
\left( \matrix{ 1 & \omega & \bar{\omega} \cr
                1 & \bar{\omega} & \omega \cr
                1 & 1 & 1 } \right) \label{tri}
\end{equation}
where $\omega=\exp(i 2 \pi /3)$ and 
$\bar{\omega}=\exp(-i2 \pi /3)$ 
are the complex cube roots of unity.
We may say that
the trimaximal mixing matrix Eq.~\ref{tri}
is the unitary matrix which 
reduces the natural representation
of $C3$ to its irreducible form.
The eigenvectors of the matrix Eq.~\ref{abc}
(appearing as the columns 
(or the complex-conjugated rows) 
of the matrix Eq.~\ref{tri})
are simply the character vectors corresponding
to the three inequivalent \mbox{(1-dimensional)}
irreducible representations of $C3$
(in which $P(\alpha \beta \gamma )$ (or $P(\gamma \beta \alpha )$)
acts like $1$, $\omega$, $\bar{\omega}$ respectively,
see Appendix A). \\

\noindent {\bf 3. The Neutrino Mass Matrix 
as an {\boldmath $S3$} Group Matrix} \vspace{2mm}

\noindent
Forced by experiment to renounce 
$C3$ invariance for leptons,
we turn with renewed interest to 
the symmetric group $S3$.
If we are to apply
the foregoing argument to $S3$,
we would expect to have to include,
in addition,
the odd $S3$ permutations:
\begin{eqnarray}
P(\alpha \beta ) = \left( 
\matrix{ 0 & 1 & 0 \cr
         1 & 0 & 0 \cr
         0 & 0 & 1 } \right) \hspace{0.5cm}  
P(\beta \gamma ) = \left( 
\matrix{ 1 & 0 & 0 \cr
         0 & 0 & 1 \cr
         0 & 1 & 0 } \right)   \hspace{0.5cm}
P(\gamma \alpha ) = \left( 
\matrix{ 0 & 0 & 1 \cr
         0 & 1 & 0 \cr
         1 & 0 & 0 } \right).  \hspace{0.5cm} \label{odd}
\end{eqnarray}
Constructing a general 
(hermitian) $S3$ group matrix
amounts to adding
an arbitrary (real) linear combination $M^2({\rm odd})$
of the odd group elements Eq.~\ref{odd} 
to the previous  
linear combination $M^2({\rm even})$
($:= M^2$ from Eq.~\ref{circ})
of the even ($C3$) operators, Eq.~\ref{even}:
\begin{eqnarray}
M^2_{\nu} & = & M^2({\rm even}) +M^2({\rm odd}) \nonumber \\ 
& = & aI +bP(\alpha \beta \gamma ) +b^*P(\gamma \beta \alpha ) +xP(\beta \gamma )+yP(\gamma \alpha )+zP(\alpha \beta ) 
\label{inter}
\end{eqnarray}
where (as above) there is nothing to be gained
by considering non-hermitian
$S3$ group matrices,   
eg.\ with $x$, $y$, $z$ complex,
which always yield the form Eq.~\ref{inter}
on taking the hermitian square 
($M \rightarrow MM^{\dagger}$).
Notice that,
within the natural representation,
the six $S3$ group operators 
(Eq.~\ref{even} together with Eq.~\ref{odd})
are {\em not} fully independent:
\begin{equation}
I+P(\alpha \beta \gamma )
          +P(\gamma \beta \alpha )
                =P(\alpha \beta )+P(\beta \gamma )
                       +P(\gamma \alpha ) \label{null}
\end{equation}
so that the effective number 
of (real) parameters 
in Eq.~\ref{inter} is 
actually only five.~\footnote{
The constraint Eq.~\ref{null}
is due to the fact that the natural representation
of the $S3$ group receives no contribution 
from the `alternating' representation
(as defined in Appendix A),
whereby the corresponding idempotent
of the $S3$ group algebra
($:=$ Eq.~\ref{null}-LHS $-$ Eq.~\ref{null}-RHS)
is identically the null matrix, 
in the natural representation.}

The structure of Eq.~\ref{inter}
may perhaps be best appreciated
by noting that the contribution
of the odd operators (from Eq.~\ref{odd})
is `retrocirculant' \cite{adler}:
\begin{equation}
M^2(\rm odd) = 
\left( \matrix{ x & z & y \cr
                z & y & x \cr
                y & x & z }  \right). \label{retroc}
\end{equation}
The significant property 
of a retrocirculant here,
is that it has 
non-degenerate eigenvalues in general,
and (clearly) always one trimaximal eigenvector
($1/\sqrt{3}$, $1/\sqrt{3}$, $1/\sqrt{3}$)!
Futhermore, it is evident
that these properties
are not in general invalidated by the inclusion
of the circulant (even) contribution
already discussed.
With $S3$ being a non-abelian group, 
Eq.~\ref{inter} 
is {\em not} invariant
under $S3$ permutations of the generation indices.
We observe that it {\em does}, however, satisfy
an $S3$ invariant constraint:
\begin{equation}
(M^2_{\nu})_{\alpha \alpha}-(M^2_{\nu})_{\beta \beta}
      =(M^2_{\nu})_{\gamma \beta}-(M^2_{\nu})_{\alpha \gamma}
\hspace{1.5cm} (\alpha \ne \beta \ne \gamma) 
\label{retro}
\end{equation}
for all ($S3$) permutations 
of the generation indices
($\alpha$, $\beta$, $\gamma$ $=$ $e$, $\mu$, $\tau$).

If Eq.~\ref{inter} is taken
to be the neutrino mass matrix
in the lepton-flavour basis
(as was already anticipated by the subscript 
on $M^2_{\nu}$ in Eq.~\ref{inter}
and by the introduction of 
explicit flavour indices
in Eq.~\ref{retro}),
then, for a suitable choice of  the coefficients, we have:
\begin{eqnarray}
m_1^2 & = & a-{\rm Re} \, b 
              - \sqrt{ \, 3 \, ({\rm Im} \, b)^2 
                                +x^2+y^2+z^2-xy-yz-zx} \\
m_2^2 & = & a + 2 {\rm Re} \, b +x+y+z \\ 
m_3^2 & = & a-{\rm Re} \, b 
              + \sqrt{ \, 3 \, ({\rm Im} \, b)^2 
                                +x^2+y^2+z^2-xy-yz-zx}
\end{eqnarray}
(`suitable' only in the sense that
the mass-eigenstates 
should turn out to be ordered appropriately, 
eg.\ $m_1^2 < m_2^2 < m_3^2$ 
for a conventional neutrino mass hierarchy).
Eq.~\ref{inter}
is seen to correspond to
the (two-parameter) mixing scheme
proposed phenomenologically in Ref.~\cite{hsym}
(to interpolate between 
tri-$\phi$maximal and tri-$\chi$maximal mixing):
\begin{eqnarray}
     \matrix{  \hspace{0.4cm} \nu_1 \hspace{1.8cm}
               & \hspace{0.5cm} \nu_2 \hspace{1.7cm}
               & \hspace{0.5cm} \nu_3  \hspace{0.5cm} }
                                      \hspace{3.5cm} \nonumber \\
\hspace{0.3cm}U \hspace{0.3cm} = \hspace{0.3cm}
\matrix{ e \hspace{0.2cm} \cr
         \mu \hspace{0.2cm} \cr
         \tau \hspace{0.2cm} }
\left( \matrix{    \sqrt{\frac{2}{3}} c_{\chi} c_{\phi} 
                 +i\sqrt{\frac{2}{3}} s_{\chi}s_{\phi}  &
                      \frac{1}{\sqrt{3}} &
                 \sqrt{\frac{2}{3}} c_{\chi}s_{\phi}
   -i\sqrt{\frac{2}{3}} s_{\chi} c_{\phi} \cr 
   -  \frac{c_{\chi}c_{\phi}-is_{\chi}s_{\phi}}{\sqrt{6}}  
     -\frac{c_{\chi}s_{\phi}+is_{\chi}c_{\phi}}{\sqrt{2}} &
          \frac{1}{\sqrt{3}} &
    \frac{c_{\chi}c_{\phi}+is_{\chi}s_{\phi}}{\sqrt{2}}  
     -\frac{c_{\chi}s_{\phi}-is_{\chi}c_{\phi}}{\sqrt{6}}  \cr
      \hspace{2mm}
     - \frac{c_{\chi}c_{\phi}-is_{\chi}s_{\phi}}{\sqrt{6}}  
     +\frac{c_{\chi}s_{\phi}+is_{\chi}c_{\phi}}{\sqrt{2}}
                                                   \hspace{2mm} &
         \hspace{2mm}
         \frac{1}{\sqrt{3}} \hspace{2mm} &
   - \frac{c_{\chi}c_{\phi}+is_{\chi}s_{\phi}}{\sqrt{2}}  
     -\frac{c_{\chi}s_{\phi}-is_{\chi}c_{\phi}}{\sqrt{6}} 
                                       \hspace{2mm} \cr } \right).
\hspace{1.5cm}
\label{int}
\end{eqnarray}
\vspace{2mm}

\noindent
In Eq.~\ref{int}
we have used the abreviations:
$c_{\chi}= \cos \chi$, $s_{\chi}= \sin \chi$,
$c_{\phi}= \cos \phi$, $s_{\phi}= \sin \phi$,
where:
\begin{eqnarray}
\tan 2 \phi & = & \frac{\sqrt{3} \, (z-y)}{z+y-2x} \\
\tan 2 \chi & = & \frac{\sqrt{3} \: {\rm Im \, b}}
               {[ \, x^2+y^2+z^2-xy-yz-zx \, ]^\frac{1}{2}} .
\end{eqnarray}
The $CP$-violation parameter $J$ \cite{jcp} 
is given by :
\begin{equation}
J= \frac{{\rm Im} \: b}{6 \, [ \, 3 \, ({\rm Im} \, b)^2 
                       +x^2+y^2+z^2-xy-yz-zx \, ]^\frac{1}{2}} 
  =\frac{\sin 2\chi}{6\sqrt{3}} .
\end{equation}
Clearly, imposing Im $b =0$ in Eq.~\ref{inter} 
(ie.\ $\chi = 0$) would imply no $CP$ violation,
whereas
imposing $y=z$ instead (ie.\ $\phi = 0$) implies
`mu-tau reflection symmetry' \cite{mutau}.
For {\em both} Im~$b =0$ and $y=z$ 
(ie.\ $\chi =0$ and $\phi =0$)
the mixing matrix Eq.~\ref{int} evidently 
reduces to the tri-bimaximal form \cite{hps5} \cite{hsym},
as does the mass matrix
(Eq.~\ref{inter}), accordingly \cite{xing}.

The six constants appearing in Eq.~\ref{inter}
may be expressed 
in terms of the three neutrino masses
and the two mixing-matrix parameters as follows:
\begin{eqnarray}
a & = & \frac{m_1^2}{3}+\frac{m_2^2}{3} + \frac{m_3^2}{3} \label{ca} \\
b & = & (-\frac{m_1^2}{6}+\frac{m_2^2}{3}-\frac{m_3^2}{6}) 
+i \; \frac{m_3^2-m_1^2}{2\sqrt{3}}\sin 2\chi \label{cb} \\
x & = & \frac{m_3^2-m_1^2}{3} \cos 2\chi \; ( \, -\cos 2 \phi \, ) \label{cx} \\
y & = & \frac{m_3^2-m_1^2}{6}
     \cos 2\chi \; ( \,  \cos 2 \phi 
                 - \sqrt{3} \sin 2 \phi \, )  
                                         \label{cy} \\
z & = & \frac{m_3^2-m_1^2}{6} 
     \cos 2\chi \; ( \, \cos 2 \phi 
                 + \sqrt{3} \sin 2 \phi \, ), 
                                         \label{cz}
\end{eqnarray}
where clearly (by virtue of Eq.~\ref{null}) 
any arbitrary constant
may be added to Eqs.\ref{ca}-\ref{cb}
provided that the same constant 
is subtracted from Eqs.~\ref{cx}-\ref{cz}.

Note that
in this approach, 
in the case of $S3$
(cf.\ the case of $C3$, Section~2)
the resulting mixing matrix (Eq.~\ref{int}) 
is {\em not} directly 
the $S3$ character table.
It is simply the 
generic unitary matrix
which reduces the natural representation of $S3$ 
to irreducible form.
The natural representation of $S3$ comprises
the trivial 1-dimensional representation
and a faithful 2-dimensional representation,
which is determined only up to 
similarity transformations
(hence the undetermined parameters
appearing in Eq.~\ref{int}). 
From its present derivation
(and to distinguish 
it clearly from mixing ansatze
based on the 
`democratic' mass matrix)
we will refer to the mixing
Eq.~\ref{int} as `$S3$ group mixing'. \\

\noindent {\bf 4. The Charged-Lepton Mass Matrix 
as an {\boldmath $S3$} Class Matrix} \vspace{2mm}

\noindent
We have seen in Section~2 that 
a very succinct way
to introduce 
trimaximal mixing 
is to demand
that one or other 
of the mass-matrices
is a $C3$ group matrix
in the natural representation of $C3$.
The mixing matrix is then
essentially
the $C3$ character table,
with all states
trimaximally mixed \cite{char}.
We went on to generalise
the argument to $S3$, 
finding that it is enough to take
one of the mass matrices
to be an $S3$ group matrix
in the natural representation of the $S3$ group,
to obtain a mixing matrix where one 
(and in particular {\em any} one)
of the eigenvectors is trimaximally mixed,
thereby `solving' the problem
of the democratic mass matrix
discussed in Section~1.

However, a more predictive 
(and perhaps more interesting) way 
to generalise the trimaximal argument 
is to recognise that, with $C3$ 
being an abelian group,
there is no distinction
between the group elements 
and the group conjugacy classes~in~that~case.
Each $C3$ group element 
being individually a class,
an arbitrary element of the $C3$ group algebra 
is also an arbitrary element  of the $C3$ class algebra. 
It is therefore not obvious
that the better generalisation to $S3$ should not
simply postulate that one or other of the  mass matrices
should live in the natural representation 
of the $S3$ class algebra,
rather than the $S3$ group algebra,
which is certainly 
a signifcantly 
different~idea. 

Following this line of thought,
we define 
(normalised) $S3$ class operators $c_i$:
\begin{eqnarray}
c_1 & = & I \label{c1} \\
c_2 & = & \frac{P(\xi \eta \zeta )
          +P(\zeta \eta \xi )}{\sqrt{2}} \label{c2}\\
c_3 & = & \frac{P(\xi \eta ) + P(\eta \zeta )
                   +P(\zeta \xi )}{\sqrt{3}} \label{c3}
\end{eqnarray}

\noindent 
where the precise physical meaning of 
$\xi$, $\eta$, $\zeta$ 
remains unclear.
Evidently,
the $S3$ class multiplication table 
(by definition commutative) 
then takes the form:
\begin{center}
\begin{tabular}{|c|c|c|c|}
\hline
        &      &     &     \\
 \hspace{2mm}  \hspace{2mm}    & \hspace{4mm} $c_1$ \hspace{4mm} &
        \hspace{4mm} $c_2$ \hspace{4mm}
                   & \hspace{4mm} $c_3$ \hspace{4mm} \\
        &      &     &     \\
\hline
     &        &          &     \\
\hspace{4mm}$c_{1}$\hspace{4mm}    &  $c_1$    & $c_2$ &    $c_{3}$ \\
    &         &          &     \\
\hspace{2mm}$c_{2}$\hspace{2mm}    
        & $c_2$     & $c_1 +c_2/\sqrt{2}$ &       $\sqrt{2}c_{3}$    \\
    &         &          &     \\
\hspace{2mm}$c_{3}$\hspace{2mm}    
         & $c_{3}$     & $\sqrt{2}c_{3}$ &       $c_1+\sqrt{2}c_2$    \\
    &         &          &     \\
\hline
\end{tabular}
\end{center}
\vspace{1mm}

\noindent 
The 
structure constants
in the table
themselves provide a 
matrix representation 
for the $c_i$
(which is the natural representation 
of the $S3$ 
class algebra 
in terms of the $c_i$):
\begin{eqnarray}
c_1 = \left( 
\matrix{ 1 & 0 & 0 \cr
         0 & 1 & 0 \cr
         0 & 0 & 1 } \right) \hspace{0.5cm}  
c_2 = \left( 
\matrix{ 0 & 1 & 0 \cr
         1 & 1/\sqrt{2} & 0 \cr
         0 & 0 & \sqrt{2} } \right)   \hspace{0.5cm}
c_3 = \left( 
\matrix{ 0 & 0 & 1 \cr
         0 & 0 & \sqrt{2} \cr
         1 & \sqrt{2} & 0 } \right)  \hspace{0.5cm} \label{crep}
\end{eqnarray}

\noindent
as is readily verified by 
direct multiplication of the matrices.

We now postulate that 
the charged-lepton mass matrix
in the neutrino mass-basis
is a suitable linear combination
of the $S3$ class operators 
in the above representation:
\begin{equation}
M^2_l = pc_1+qc_2+rc_3, \label{ml2}
\end{equation}
ie.\ explicitly:
\begin{equation}
      M_l^2 =  \left( \matrix{
p & q & r \cr
q & p+q/\sqrt{2} & \sqrt{2}r \cr
r & \sqrt{2}r & p+\sqrt{2}q } \right). \label{pqr}  
\end{equation}
From the usual argument (see eg.\ Section 2) 
the coefficients 
$p$, $q$, $r$ may be taken to be real.
The eigenvalues of the matrix Eq.~\ref{pqr}
are then the charged-lepton masses:
\begin{eqnarray}
m_e^2 & = & p-q/\sqrt{2} \label{me} \\
m_{\mu}^2 & = & p+\sqrt{2}q-\sqrt{3}r \label{mm} \\
m_{\tau}^2 & = & p+\sqrt{2}q+\sqrt{3}r. \label{mt}
\end{eqnarray}
The coefficients
(being `suitable' only 
in that $0<r/\sqrt{3}<q/\sqrt{2}<p$
to order the mass-eigenstates
in accord with experiment) 
are expressible 
in terms of the masses by:
\begin{eqnarray}
p & = & \frac{m_{\tau}^2+m_{\mu}^2}{6}+\frac{2}{3}m_e^2 \\
q & = & \sqrt{2} ( \frac{m_{\tau}^2+m_{\mu}^2}{6}-\frac{m_e^2}{3}) \\
r & = & \frac{m_{\tau}^2-m_{\mu}^2}{2\sqrt{3}}.
\end{eqnarray}
The unitary matrix diagonalising Eq.~\ref{pqr}
(independent of the values of $p$, $q$ and $r$) is 
directly the tri-bimaximal mixing matrix:
\begin{eqnarray}
     \matrix{  \hspace{0.4cm} \nu_1 \hspace{0.2cm}
               & \hspace{0.2cm} \nu_2 \hspace{0.3cm}
               & \hspace{0.2cm} \nu_3  \hspace{0.4cm} }
                                      \hspace{1.2cm} \nonumber \\
\hspace{1.0cm}U \hspace{0.3cm} = \hspace{0.3cm}
\matrix{ e \hspace{0.2cm} \cr
         \mu \hspace{0.2cm} \cr
         \tau \hspace{0.2cm} }
\left( \matrix{ \frac{2}{\sqrt{6}}  &
                     -\frac{1}{\sqrt{3}} &
                                   0  \cr
 \frac{1}{\sqrt{6}}  &
          \frac{1}{\sqrt{3}} &
 -\frac{1}{\sqrt{2}}   \cr
      \hspace{2mm}
  \frac{1}{\sqrt{6}} \hspace{2mm} &
         \hspace{2mm}
            \frac{1}{\sqrt{3}} \hspace{2mm} &
                                 \frac{1}{\sqrt{2}}
                                       \hspace{2mm} \cr } \right)
\label{tbm}
\hspace{1.0cm}
\label{tbx}
\end{eqnarray}
(the neutrino mass-eigenstates
having been already implicitly
ordered in accord with experiment,
by the labelling  
of the class operators,
in Eqs.~\ref{c1}-\ref{c3}).
The charged-lepton mass eigenstates
(ie.\ the eigenvectors of Eq.~\ref{pqr})
appear as the rows of Eq.~\ref{tbm}.

Clearly, 
the tri-bimaximal mixing matrix
Eq.~\ref{tbm} is very closely related 
to the $S3$ table of characters
(cf.\ as displayed in Appendix A below).
In fact, 
it differs only by  
the class-dependent normalisation factors
introduced into Eqs.~\ref{c1}-\ref{c3}.
Diagonalising 
an $S3$ class matrix (such as Eq.~\ref{pqr})
is entirely equivalent \cite{chen} 
to determining the $S3$ group characters,
ie.\ to finding all 
the irreducible represenations of $S3$
by reducing the $S3$ class algebra 
to independent idempotents.
Explicitly:
\begin{equation}
U_{li}= \sqrt{\frac{g_i}{g}}\chi^{(l)}_{i} \label{uchi}
\end{equation} 
where
$\chi^{(l)}_{i}$ is the 
$i$-th component of 
the $l$-th character vector,
$g_{i}$ is the order of the class
($g_{i}=1,2,3$ for $i=1,2,3$ for $S3$)
and $g$ is the order of the group
($g=6$ for $S3$).
Individual character components 
are simply related \cite{chen}
to the eigenvalues of the corresponding class operators,
whereby the charged-lepton masses may also be expressed
(equivalently to Eqs.~\ref{me}-\ref{mt})
in terms of the $S3$ group characters
and the constants $(p_1,p_2,p_3):=(p,q,r)$,
as follows:
\begin{equation}
m^2_l= \sum_{i} p_{i} \frac{\sqrt{g_{i}}}{d_l}\chi^{(l)}_{i}
\end{equation} 
where $d_l$ is the dimension of the irreducible
representation corresponding to the lepton flavour $l$.
The irreducible representations 
for $l=\tau$ and $l=\mu$ are the two mutually
conjugate 1-dimensional representations
(the trivial and alternating
representations respectively),
while the electron ($l=e$) is to be associated 
with the 2-dimensional faithful representation
having a self-conjugate tableau.
In the extreme hierarchical
limit,~$ r/\sqrt{3} \rightarrow q/\sqrt{2} \rightarrow p \: $
in Eqs.~\ref{me}-\ref{mt},
only the trivial representation has mass.

It is perhaps worth  re-iterating 
at this point that 
`data on neutrino oscillations point strongly \dots 
to tri-bimaximal mixing' \cite{hps5}.
We note that from its present derivation,
and in view of the need to distinguish it from `$S3$ group mixing'
(Section~3), tri-bimaximal mixing
might reasonably be termed 
`$S3$ class mixing'.\\

\noindent
{\bf 5. The Neutrino Mass Matrix
as an {\boldmath $S3 \supset S2$} 
Class Operator}. \vspace{3mm}

\noindent
Alerted to the relevance
of class operators,
we may now return 
to reconsider
the neutrino mass matrix
in the flavour basis.
According to Section~3,
the charged-lepton flavour basis
($\alpha$, $\beta$, $\gamma$ = $e$, $\mu$, $\tau$)
carries the natural representation
of the $S3$ group
(it was also noted that tri-bimaximal mixing 
requires a particular $S3$ group matrix 
with Im~$b=0$ and $y=z$).
Clearly any representation of a group 
also provides a representation for the classes,
and seeking consistency
with the results of Section~4,
we now postulate that the
neutrino mass matrix in the flavour basis
is a class operator 
for the canonical subgroup chain
$S1 \subset S2 \subset S3$
in the natural
representation of the $S3$ group
(class operators for successive 
subgroups clearly commute).
The individual class operators
may be written:
\begin{eqnarray}
C(1) = I = \left( 
\matrix{ 1 & 0 & 0 \cr
         0 & 1 & 0 \cr
         0 & 0 & 1 } \right) \hspace{15mm}
C(2) = P(\mu \tau)  = \left( 
\matrix{ 1 & 0 & 0 \cr
         0 & 0 & 1 \cr
         0 & 1 & 0 } \right) \\ 
C(3) = P(e \mu) + P(\mu \tau)+P(\tau e)   
= \left( 
\matrix{ 1 & 1 & 1 \cr
         1 & 1 & 1 \cr
         1 & 1 & 1 } \right) \hspace{27mm}
\end{eqnarray}
(class normalisation factors
are not needed here
since they may be absorbed into 
the coefficients $s$, $t$, $u$ below,
with no change of basis involved).
The $S2$ class operator $C(2)$
has been chosen to be the $\mu-\tau$ 
interchange operator \cite{mutau}.
Of course, $C(3)$ is familiar as the 
`democratic' mass matrix.

The most general (hermitian)
$S1 \subset S2 \subset S3$
class operator may be written:
\begin{equation}
M^2_{\nu}  =  sC(1)+tC(2)+uC(3). \label{s3s2}
\end{equation}
Explicitly:
\begin{equation}
M^2_{\nu}  =  \left( \matrix{
s+t+u & u & u \cr
u   & s+u & t+u \cr
u & t+u & s+u } \right) \label{m3s2}
\end{equation}
where $s$, $t$, $u$ are real.
The eigenvalues 
of the matrix Eq.~\ref{s3s2}
are the neutrino masses:
\begin{eqnarray}
m_1^2 & = & s+t \\
m_2^2 & = & s+t+3u \\
m_3^2 & = & s-t.
\end{eqnarray}
The coefficients
$0 \le 3u \le -2t \le 2s$
(for $m_1^2 \le m_2^2 \le m_3^2$) 
are given in terms of
the neutrino masses by:
\begin{eqnarray}
s & = & \frac{m_1^2+m_3^2}{2} \\
t & = & \frac{m_1^2-m_3^2}{2} \\
u & = & \frac{m_2^2-m_1^2}{3}
\end{eqnarray}
now with no arbitrary constant involved 
(cf.\ Eqs.~\ref{ca}-\ref{cz}).
The extreme hierarchical limit 
for the neutrino masses is approached 
as $u \rightarrow 0$ and $t \rightarrow -s$, 
when only the $\nu_3$ has mass. 
It may be remarked that it is
the `democratic' component $C(3)$
which has the (numerically) smallest coefficient ($u$)
in Eq.~\ref{s3s2},
vanishing in the hierarchical limit.

Diagonalising the mass matrix Eq.~\ref{m3s2},
the resulting mixing matrix 
takes the familiar 
(Eq.~\ref{tb0}) tri-bimaximal form,
which is also referred to here as `$S3 \supset S2$ mixing': \vspace{-7mm}
\begin{eqnarray}
     \matrix{  \hspace{0.4cm} \nu_1 \hspace{0.3cm}
               & \hspace{0.2cm} \nu_2 \hspace{0.3cm}
               & \hspace{0.2cm} \nu_3  \hspace{0.4cm} }
                                      \hspace{0.2cm} \nonumber \\
\hspace{1.0cm}U \hspace{0.3cm} = \hspace{0.3cm}
\matrix{ e \hspace{0.2cm} \cr
         \mu \hspace{0.2cm} \cr
         \tau \hspace{0.2cm} }
\left( \matrix{ \sqrt{\frac{2}{3}}  &
                     \frac{1}{\sqrt{3}} &
                                   0  \cr
 -\frac{1}{\sqrt{6}}  &
          \frac{1}{\sqrt{3}} &
 -\frac{1}{\sqrt{2}}   \cr
      \hspace{2mm}
 -\frac{1}{\sqrt{6}} \hspace{2mm} &
         \hspace{2mm}
            \frac{1}{\sqrt{3}} \hspace{2mm} &
                                 \frac{1}{\sqrt{2}}
                                       \hspace{2mm} \cr } \right).
\label{tb2}
\end{eqnarray}
The eigenvectors of Eq.~\ref{m3s2}
appear as the columns of Eq.~\ref{tb2}.

Clearly, the mass matrix Eq.~\ref{m3s2}
may equally well be viewed as the particular
\mbox{`$S3$ group matrix'} \cite{little}
having ${\rm Im} \: b = 0$ and $y=z$ (see Section~3).
The $\nu_2$ has the trimaximal eigenvector 
\mbox{($1/\sqrt{3}$, $1/\sqrt{3}$, $1/\sqrt{3}$)}
which is in effect the character vector
of the trivial 1-dimensional (symmetric) 
representation of $S3$.
The $\nu_1$ and $\nu_3$
both belong to the self-conjugate 
(faithful) \mbox{2-dimensional} representation of $S3$,
being distinguisehd here 
by their symmetry ($\pm 1$ respectively) under 
$\mu-\tau$ exchange 
(`mutativity'~\cite{mutau}).
The tri-bimaximal mixing matrix 
is then properly regarded
as nothing more than the table of induction coefficients
for the $[2] \otimes [1] = [3] + [21]$
induced representation of $S3$.
It is simply the unitary matrix 
which reduces the natural representation of $S3$,
whilst simultaneously diagonalising
the $\mu-\tau$ interchange operator \cite{mutau}.

In retrospect, 
the original circulant mass matrix \cite{hs1}
leading to trimaximal mixing
might have been seen as a class operator
for the group chain $S3 \supset C3$.
Of course we now know that,
for the neutrino mass matrix in the flavour basis,
an $S3 \supset S2$ class operator Eq.~\ref{s3s2},
is preferred experimentally
(the `$ \supset S1$' in fact carries 
no additional symmetry information
and is dropped here 
in accord with usual practice). \\

\noindent {\bf 6. Discussion} \vspace{2mm}

\noindent
We have been 
to a large degree logically led,
from the original trimaximal hypothesis,
first to `$S3$ group mixing' Eq.~\ref{int},
and then on to `$S3$ class mixing'
or `$S3 \supset S2$ mixing',
ie.\ to tri-bimaximal mixing.   
The two levels of generalisation 
are not inconsistent:
the latter is clearly more restrictive,
in that exact tri-bimaximal mixing
{\em requires} the charged-lepton
mass matrix to be an $S3$ class matrix
in the neutrino mass-basis,
and {\em also} requires 
the neutrino mass matrix
in the lepton flavour basis
to be a particular $S3$ group matrix
(with Im~$b=0$ and $y=z$), 
ie.\ an $S3 \supset S2$ class operator.
For a discussion 
of the forms
of both mass-matrices
in an intermediate basis
see Ref.~\cite{hsym}.

Thus, while `$S3$ group mixing'
is regarded as an interesting
mixing ansatz in its own right \cite{hsym},
our main results relate to tri-bimaximal mixing, 
and the link 
to the $S3$ group characters \cite{little}
via Eq.~\ref{uchi} (Section~4)
and to the $S3$ induction coefficients \cite{chen}
(Section~5).
In the first case
the neutrino mass eigenstates 
are associated
with the normalised $S3$ class operators
Eqs.~\ref{c1}-\ref{c3} ($\nu_i \sim c_i$),
while the charged-lepton mass-eigenstates
are in correspondence
with the $S3$ irreducible representations.
Then, in the flavour basis,
the charged leptons $e$, $\mu$, $\tau$ are
in correspondence with the $C3$ classes 
$c_0$, $c_-$, $c_+$ respectively
(viewed as the coset representatives 
with respect to the $\mu-\tau$ exchange subgroup)
while the neutrino mass-eigenstates
are in correspondence with the 
irreducible basis vectors
of the corresponding
induced representation of $S3$.
Clearly classes
(and hence linear combinations of classes)
are always permutation invariants.

Finally, 
we remark that
the notion of the yukawa couplings here
being related to the structure constants
of a permutation class algebra,
is not so different in character
from the established notion
of the couplings between gauge bosons
being the structure constants
of a lie algebra.
Of course as always, experiment 
will be the ultimate judge,
with the detailed experimental predictions
of exact tri-bimaximal mixing
(eg.\ $P(e \rightarrow e) \rightarrow 5/9 \simeq 0.56$ 
in KAMLAND \cite{kaml}, 
zero $CP$ violation,
no high-energy matter resonance etc.)
being already documented
in the literature \cite{hps5}.

\vspace{5mm}
\noindent {\bf Acknowledgement}

\noindent
We thank P.\ Slodowy for helpful 
explanations on group characters. 
This work was supported by the UK
Particle Physics and Astronomy Research Council
(PPARC). \vspace{6mm}

\noindent {\bf Appendix A: Group Character Tables 
for the {\boldmath $C3$ and $S3$} Groups } \vspace{1mm}

\noindent
For ease of reference, 
the character tables for
the cyclic group $C3$ on three symbols, 
and for the corresponding symmetric group $S3$,
are given below.  

For $C3$
there are three irreducible representations,
all \mbox{1-dimensional,}
where the generator of (say `clockwise')
cyclic permutations acts like
$1$, $\omega$ or $\bar{\omega}$,
which are referred to here as 
the trivial, $\omega$
and $\bar{\omega}$-representations respectively.
The three classes ($c_0, c_+, c_-$) 
comprise the identity, clockwise
and anti-clockwise cyclic permutations, respectively.

For $S3$
there are likewise three irreducible representations,
two of which are \mbox{1-dim}-ensional.
In the trivial representation all 
group elements act like (+1), while
in the alternating representation,
elements corresponding to
odd permutations act instead like (-1).
There is a faithful 2-dimensional representation
which may be written~\cite{ljan}:
\[
 I = \left( 
\matrix{ 1 & 0 \cr
         0 & 1 } \right)  \hspace{2.1cm} \vspace{-1mm}
\]
\[
 P(\alpha \beta \gamma ) = \left( 
\matrix{ -1/2 & \sqrt{3}/2 \cr
         -\sqrt{3}/2 & -1/2 } \right)   \hspace{0.5cm}
P(\gamma \beta \alpha ) = \left( 
\matrix{ -1/2 & -\sqrt{3}/2 \cr
         \sqrt{3}/2 & -1/2 } \right)   \hspace{1.2cm}  
                                        \vspace{1mm}
\]
\[
P(\alpha \beta ) = \left( 
\matrix{ -1/2 & \sqrt{3}/2 \cr
         \sqrt{3}/2 & 1/2 } \right) \hspace{0.3cm}  
P(\beta \gamma ) = \left( 
\matrix{ 1 & 0 \cr
         0 & -1 } \right)    \hspace{0.3cm}
P(\gamma \alpha ) = \left( 
\matrix{ -1/2 & -\sqrt{3}/2 \cr
         -\sqrt{3}/2 & 1/2 } \right)  \hspace{0.5cm}
\] 
(up to equivalence transformations).
The three classes ($c_i$, $i=1-3$) correspond 
to the identity, the even (ie.\ cyclic),
and odd permutations, respectively.

The group character tables below
give the traces $\chi_{i}^{(l)}$
of the matrices comprising 
all the inequivalent 
irreducible representations $(l)$
of the group,
as a function of conjugacy class $c_i$.
(Matrices representing different 
group elements within the same class
have the same trace,
and traces are unaltered 
by equivalence transformations).

\vspace{0.5cm}

\begin{center}
\begin{tabular}{|c|c|c|c|}
\hline
\hline
        & \multicolumn{3}{c|} { }  \\
 Cyclic Group $C3$       & \multicolumn{3}{c|} 
                           {Conjugacy Classes $\rightarrow$} \\
  \hspace{1mm} (order of group: $g = 3$) \hspace{1mm}  
   & \multicolumn{3}{c|} {(order of class, $g_{i}$)} \\
        & \multicolumn{3}{c|} { } \\
\hline
        &      &     &     \\
 \hspace{2mm} Irreducible \hspace{2mm}    & \hspace{4mm} $c_0$ \hspace{4mm} &
        \hspace{4mm} $c_+$ \hspace{4mm}
                   & \hspace{4mm} $c_-$ \hspace{4mm} \\
 \hspace{2mm} Representations $\downarrow$ \hspace{2mm}    
             & \hspace{4mm} $(1)$ \hspace{4mm} &
        \hspace{4mm} $(1)$ \hspace{4mm}
                   & \hspace{4mm} $(1)$ \hspace{4mm} \\
        &      &     &     \\
\hline
\hline
     &        &          &     \\
$\omega$-rep.\    &  $1$    & $\omega$ &    $\bar{\omega}$ \\
    &         &          &     \\
$\bar{\omega}$-rep.\    & $1$     & $\bar{\omega}$ &       $\omega$    \\
    &         &          &     \\
triv.\    & $1$     & $1$ &       $1$    \\
    &         &          &     \\
\hline
\hline
\end{tabular}
\end{center}

\vspace{0.1cm}
\begin{center}
\begin{tabular}{|c|c|c|c|}
\hline
\hline
        & \multicolumn{3}{c|} { }  \\
 Symmetric Group $S3$       & \multicolumn{3}{c|} 
                           {Conjugacy Classes $\rightarrow$} \\
  \hspace{1mm} (order of group: $g = 6$) \hspace{1mm}  
   & \multicolumn{3}{c|} {(order of class, $g_{i}$)} \\
        & \multicolumn{3}{c|} { } \\
\hline
        &      &     &     \\
 \hspace{2mm} Irreducible \hspace{2mm}    & \hspace{4mm} $c_1$ \hspace{4mm} &
        \hspace{4mm} $c_2$ \hspace{4mm}
                   & \hspace{4mm} $c_3$ \hspace{4mm} \\
 \hspace{2mm} Representations $\downarrow$ \hspace{2mm}    
             & \hspace{4mm} $(1)$ \hspace{4mm} &
        \hspace{4mm} $(2)$ \hspace{4mm}
                   & \hspace{4mm} $(3)$ \hspace{4mm} \\
        &      &     &     \\
\hline
\hline
     &        &          &     \\
faith.\    &  $2$    & $-1$ &    $0$ \\
    &         &          &     \\
alt.\    & $1$     & $1$ &       $-1$    \\
    &         &          &     \\
triv.\    & $1$     & $1$ &       $1$    \\
    &         &          &     \\
\hline
\hline
\end{tabular}
\end{center}

\noindent
The following abbreviations have been used:
$\omega$-rep.\ = $\omega$-representation, 
triv.\ = trivial, 
alt.\ = alternating
and faith.\ = faithful (representations).  
The complex cube roots of unity are given by:
$\omega=\exp(i2\pi/3)$ and $\bar{\omega}=\exp(-i2\pi/3)$.

\end{document}